\documentclass{article}
\usepackage{graphicx}
\usepackage{amsmath}
\usepackage[top=1in, bottom=1in,left=1in,right=1in]{geometry}
\usepackage{xcolor}
\usepackage{float}

\begin{document}
\title{Dark boson mediation of the $\pi^0\rightarrow\gamma e^+e^-$ decay}%
\author{M. Naydenov$^1$, V. Kozhuharov$^1$}%
\date{}
\maketitle
\begin{center}
$^1${Faculty of Physics, Sofia University, 5 J. Bourchier Blvd, 1164 Sofia, Bulgaria}
\end{center}
\flushbottom

\section*{Abstract}
The phenomenological implications to the decay of the neutral pion 
from the introduction of new dark particles are discussed. 
We calculate the
contribution to the 
$\pi^0$ 
decay width and then we extend the theory by adding dark sector vector particles which interact through $\gamma^\mu$, $\sigma^{\mu\nu}$ and $\sigma^{\mu\nu}\gamma^5$. We calculate the total decay width of $\pi^0\rightarrow\gamma e^+e^-$ where the electron-positron pair is produced through a decay of a dark meson. 
The implication of the presented model on the present and future searches 
for new dark particles is also discussed.

\section{Introduction}

The existence of dark photons was theoretically proposed within the framework of quantum field theory in 1986 by B. Holdom \cite{holdom}. He discusses the possibility of having an additional U$(1)$ gauge theory in analogy with electrodynamics and the constant of interaction to be $e'=\epsilon e,$ where $\epsilon$ is some small rescaling factor restricting the intense interaction with visible matter. In this sense the photon becomes a mixed state between two separate vector states. Many experiments have been performed at colliders to find a dark particle, more recent reviews on the topic are \cite{rev1,rev2,rev3}.

The $^8$Be anomalous decay \cite{Be} and the muon magnetic moment anomaly \cite{g2} are currently considered as the main motivation for searching for dark sector physics although these peculiarities find their explanation within the Standard Model \cite{SM1,SM2}. In our work we present a different perspective to these problems by extending the existing U$(1)$ model. From the $^8$Be decay we know that if it is due to a dark sector particle it has to have a mass around 17 MeV/c$^2$ and is most probably a vector meson, or a pseudoscalar (highly suppressed). But even with the existence of a dark vector particle, or a dark massive photon, the discrepancy between the theoretically and the experimentally obtained value for the magnetic moment of the muon is not completely compensated. There also exist various explanations for the observed phenomenon such as the existence of leptoquarks \cite{leptoquarks} or effects from supersymmetry \cite{supersymmetry}.

In our work we will focus on the phenomenology that hidden sector particles could have on the electromagnetic decays of the neutral pion. The predominant decay mode is $\pi^0\rightarrow\gamma\gamma$ with a relative fraction $\Gamma_1=(98.823\pm0.034)\%$ \cite{pdg}. We investigate the missing energy case when we observe just a single photon, instead of two. Extensive discussions on the phenomenology of $\pi^0$ dark decays and potential channels for production are \cite{photon1,photon2,photon3,photon4}. Effective Lagrangian describing the strong and electromagnetic interactions governing the pion decay can be constructed for the case of a dark photon.
 
Within the Nambu--Jona-Lasinio model the full phenomenological description of the $\sigma$, $\pi$, $\rho$, $a_1$, $\rho'$ and $b_1$ mesons can only be achieved with the inclusion of tensor currents \cite{chizhov1,chizhov_naydenov,osipov}. One of the main obstacles in introducing tensor interactions is that they are equivalently equal to 0 if chiral symmetry is to be obeyed \cite{Eguchi}. We can extend the existing Lagrangian in a non-trivial manner to include tensor currents by using $\sigma_{\mu\nu}$ and $\sigma_{\mu\nu}\gamma^5$. Stepping on the successful models for the low lying meson states we explore the consequences on the aforementioned anomalies of having two additional dark sector vector fields which interact through tensors. Such particles give a contribution to the pion decay and the production of an electron-positron pair can be observed. We are specifically interested in the influence of dark sector particles to the decay $\pi^0\rightarrow\gamma e^+e^-$ where the electron-positron pair is mediated by a dark sector particle. 

The outline of our work is as follows: we consider the neutral pion decay into a pair of photons and introduce dark sector particles. After we have the widths for dark meson decays into an electron-positron pair we can account for the total contribution to the Dalitz decay from dark mesons. Assuming that the experimental error is due to influence of interactions beyond the Standard Model we can set a limit on the values that the coupling constant $\epsilon e$ can obtain. 

We also consider several benchmark points for the dark boson mass, namely 5 MeV, 17 MeV, 50 MeV, and 100 MeV. 

\section{Phenomenology of $\pi^0\rightarrow\gamma\gamma$ and $\pi^0\rightarrow\gamma\gamma'$}

\begin{figure}[H]
\centering
\includegraphics[width = 0.46\textwidth]{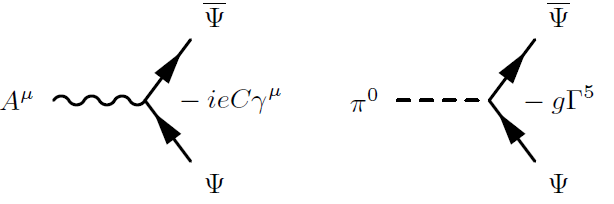}
 \caption{Basic Feynman diagrams for the pion and the photon in this model.}
 \label{FRules}
\end{figure}

We start with an effective Lagrangian describing the strong and electromagnetic interactions which govern the decay of $\pi^0$ into two photons:
\begin{equation}
\begin{gathered}
\mathcal{L}=\mathcal{L}_{\text{strong}}+\mathcal{L}_{\text{EM}}=ig\bar{\Psi}\gamma^5\vec{\tau}\Psi\vec{\pi}-e\left(\frac{2}{3}\bar{u}\gamma_\mu u-\frac{1}{3}\bar{d}\gamma_\mu d\right)A^\mu-m\overline{\Psi}\Psi,
\end{gathered}
\label{pion_L}
\end{equation}
where
 \begin{align}
    \Psi &= \begin{pmatrix}
          u \\
          d
         \end{pmatrix},
          \hspace{10mm}
    \vec{\pi}=\begin{pmatrix}
    	\pi_1\\
    	\pi_2\\
    	\pi_3
    \end{pmatrix},
  \end{align}
\noindent and $g$ is the pion decay constant. Written in a more compact form 

\begin{equation}
\mathcal{L}=ig\bar{\Psi}\Gamma^5\Psi\pi^0-e\bar{\Psi}C\hat{A}\Psi-m\overline{\Psi}\Psi,
\end{equation}
\noindent where $\hat{A}=\gamma_\mu A^\mu$, $\Gamma^5=\gamma^5\tau^3$ and $C=\frac{1}{6}\tau^0+\frac{1}{2}\tau^3$. The emerging Feynman rules for this model are shown in 
fig. \ref{FRules}.

\begin{figure}[H]
\centering
\includegraphics[width = 0.46\textwidth]{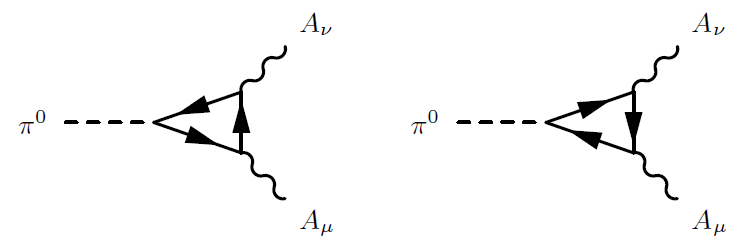}
\caption{Feynman diagram for the decay of a neutral pion into two photons.}
\label{fig:pi-ggp}
\end{figure}

The matrix element of the decay can be deduced by applying the Feynman rules to the diagrams in 
fig. \ref{fig:pi-ggp}.

\begin{equation}
\begin{gathered}
\mathcal{M}_{\pi^0\gamma\gamma}^{\mu\nu}=-ige^2 \int\frac{d^4p}{(2\pi)^4}\Bigg\{\frac{\text{Tr}\left[\gamma^5(\hat{p}+m)\gamma^\mu(\hat{p}+\hat{k}+m)\gamma^\nu(\hat{p}+\hat{q}+m)\right]}{(p^2-m^2)[(p+k)^2-m^2][(p+q)^2-m^2]}\\
+\frac{\text{Tr}\left[\gamma^5(\hat{p}-\hat{q}+m)\gamma^\nu(\hat{p}-\hat{k}+m)\gamma^\mu(\hat{p}+m)\right]}{(p^2-m^2)[(p-k)^2-m^2][(p-q)^2-m^2]}\Bigg\}=\\
=-8 ge^2m\epsilon^{qk\mu\nu}\int\frac{d^4p}{(2\pi)^4}\frac{1}{(p^2-m^2)[(p+k)^2-m^2][(p+q)^2-m^2]}.
\end{gathered}
\end{equation}

\noindent This is a convergent integral and can be solved by applying the Feynman parameters method, which within the chiral limit (massless pion) equates to

\begin{equation}
\begin{gathered}
\int\frac{d^4p}{(2\pi)^4}\frac{1}{(p^2-m^2)[(p+k)^2-m^2][(p+q)^2-m^2]}=\\
-\frac{i}{16\pi^2m_\pi^2}\left[\text{Li}_2\left(\frac{2m_\pi}{m_\pi-\sqrt{m_\pi^2-4m^2}}\right)+\text{Li}_2\left(\frac{2m_\pi}{m_\pi+\sqrt{m_\pi^2-4m^2}}\right)\right]\\
\rightarrow -\frac{i}{32\pi^2m^2}\text{ for }m_\pi\rightarrow 0.
\end{gathered}
\end{equation}

\noindent The Goldberger-Treiman relation gives us the ratio between the constant $g$ and the constituent quark mass $m$:

 \begin{equation}
g=\frac{m g_A}{f_\pi},\text{ where we take }f_\pi=92.1\pm0.3 \text{ MeV}/c^2~\cite{decay_constant} \text{ for  point particles } g_A=1~\cite{weinberg}.
\end{equation}

\noindent The matrix element in the simplest possible form is

\begin{equation}
\mathcal{M}_{\pi^0\gamma\gamma}^{\mu\nu}=i\frac{e^2}{4\pi^2 f_\pi}\epsilon^{\mu\nu\alpha\beta}q_{\alpha}k_{\beta}.
\end{equation}
\noindent Working in natural units $e=0.303$, $m=336\text{ MeV}/\text{c}^2$ \cite{Griffiths}, $m_\pi=134.9768\pm0.0005\text{ MeV}/\text{c}^2$ \cite{pdg2} the decay rate is

\begin{equation}
\Gamma=\frac{1}{2m_\pi}\frac{1}{8\pi}\frac{1}{2}|M|^2=\frac{1}{32\pi m_\pi}\left(\frac{e^2}{4\pi^2 f_\pi}\right)^22(qk)^2=\frac{e^4 m_\pi^3}{1024\pi^5 f_\pi^2} \approx 7.8 eV,
\end{equation}

\noindent where an additional factor of $\frac{1}{2}$ is added because of the indistinguishability of the two photons. 
Substituting all known values we obtain that the lifetime of the neutral pion is about $\tau=8.4157\times 10^{-17}$s.

The described procedure can also be applied to calculate the 
decay rate for the $\eta\rightarrow\gamma\gamma$ decay. 
While the result 
\begin{equation}
    \Gamma (\eta\rightarrow\gamma\gamma) =\frac{m_{\eta}^3}{64\pi}\left(\frac{\alpha}{ \sqrt{3}\pi f_\pi}\right)^2  \approx 173 eV
\end{equation}
reproduces the result in \cite{meson_decays}, both differ from the experimental value
$\Gamma (\eta\rightarrow\gamma\gamma) = (516 \pm 18)$ eV \cite{pdg} due to the effect of $\eta-\eta'$ mixing and higher order contributions. 

We extend the Lagrangian (\ref{pion_L}) by adding a term for the interaction with a dark photon particle

\begin{equation}
\mathcal{L}=ig\overline{\Psi}\Gamma^5\Psi\pi^0-e\overline{\Psi}C\gamma_\mu A^\mu\Psi-e_1\overline{\Psi}C\gamma_\mu A_1^\mu\Psi-m\overline{\Psi}\Psi,
\end{equation}

\noindent where $A_1^\mu$ is the dark photon field and $e_1$ is the constant of interaction defined as $e_1=\epsilon e$ 
\cite{DP_Production}. 
Then the pion decay is calculated applying the Feynman rules to the diagrams
\vspace{5mm}

\begin{figure}[H]
\centering
\includegraphics[width = 0.46\textwidth]{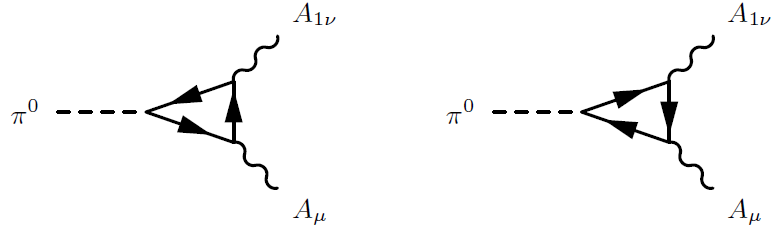}
\caption{Feynman diagrams for the decay of a neutral pion into a photon and a dark photon.}
\end{figure}

\noindent Assuming the existence of three colours after applying the Feynman rules for the theory the effective Lagrangian becomes
\begin{equation}
\begin{gathered}
\mathcal{M}_{\pi^0 \gamma\gamma'}^{\mu\nu}=ee_1g\int\frac{d^4p}{(2\pi)^4}\frac{\text{Tr}\left[\gamma^5(\hat{p}-\hat{q}+m)\gamma^\nu(\hat{p}-\hat{k}+m)\gamma^\mu(\hat{p}+m)\right]}{(p^2-m^2)\left[(p-k)^2-m^2\right]\left[(p-q)^2-m^2\right]}+\\
+ee_1g\int\frac{d^4p}{(2\pi)^4}\frac{\text{Tr}\left[\gamma^5(\hat{p}+m)\gamma^\mu(\hat{p}+\hat{k}+m)\gamma^\nu(\hat{p}+\hat{q}+m)\right]}{(p^2-m^2)\left[(p+k)^2-m^2\right]\left[(p+q)^2-m^2\right]}.
\end{gathered}
\end{equation}

\noindent After simplification we get

\begin{equation}
\begin{gathered}
|\mathcal{M}_{\pi^0 \gamma\gamma'}^{\mu\nu}|^2=\frac{1}{4\pi^4}m^2e^2e_1^2g^2\left(\frac{1}{2m^2}+\frac{m_1^2}{24m^4}\right)^2\epsilon^{\mu\nu\alpha\beta}\epsilon^{\mu\nu\sigma\rho}q^\alpha k^\beta q^\sigma k^\rho=\\
\frac{m^2e^2e_1^2g^2}{8\pi^4}\left(\frac{1}{2m^2}+\frac{m_1^2}{24m^4}\right)^2(m_\pi^2-m_1^2)^2,
\end{gathered}
\end{equation}

\noindent where $m_1$ is the dark photon mass. The decay width of this process becomes
\begin{equation}
\begin{gathered}
\Gamma_{\pi^0\gamma\gamma'}=\frac{1}{16\pi m_\pi^3}(m_\pi^2-m_1^2)|\mathcal{M}|^2=\\
=\frac{1}{128\pi^5 m_\pi^3}(m_\pi^2-m_1^2)^3 \epsilon^2e^4\frac{m^4}{f_\pi^2}\left(\frac{1}{2m^2}+\frac{m_1^2}{24m^4}\right)^2.
\end{gathered}
\label{decay_width}
\end{equation}
For $\epsilon = 10^{-2}$ and $m_1 = 17$ MeV this corresponds to $\Gamma_{\pi^0\gamma\gamma'}=1.43 \text{ meV} $ or a branching ratio of $B_f=18.7\times 10^{-5}$. The branching ratio can be plotted against the dark photon mass, which is shown in the figure below.
\begin{figure}[H]
\centering
\includegraphics[scale=0.6]{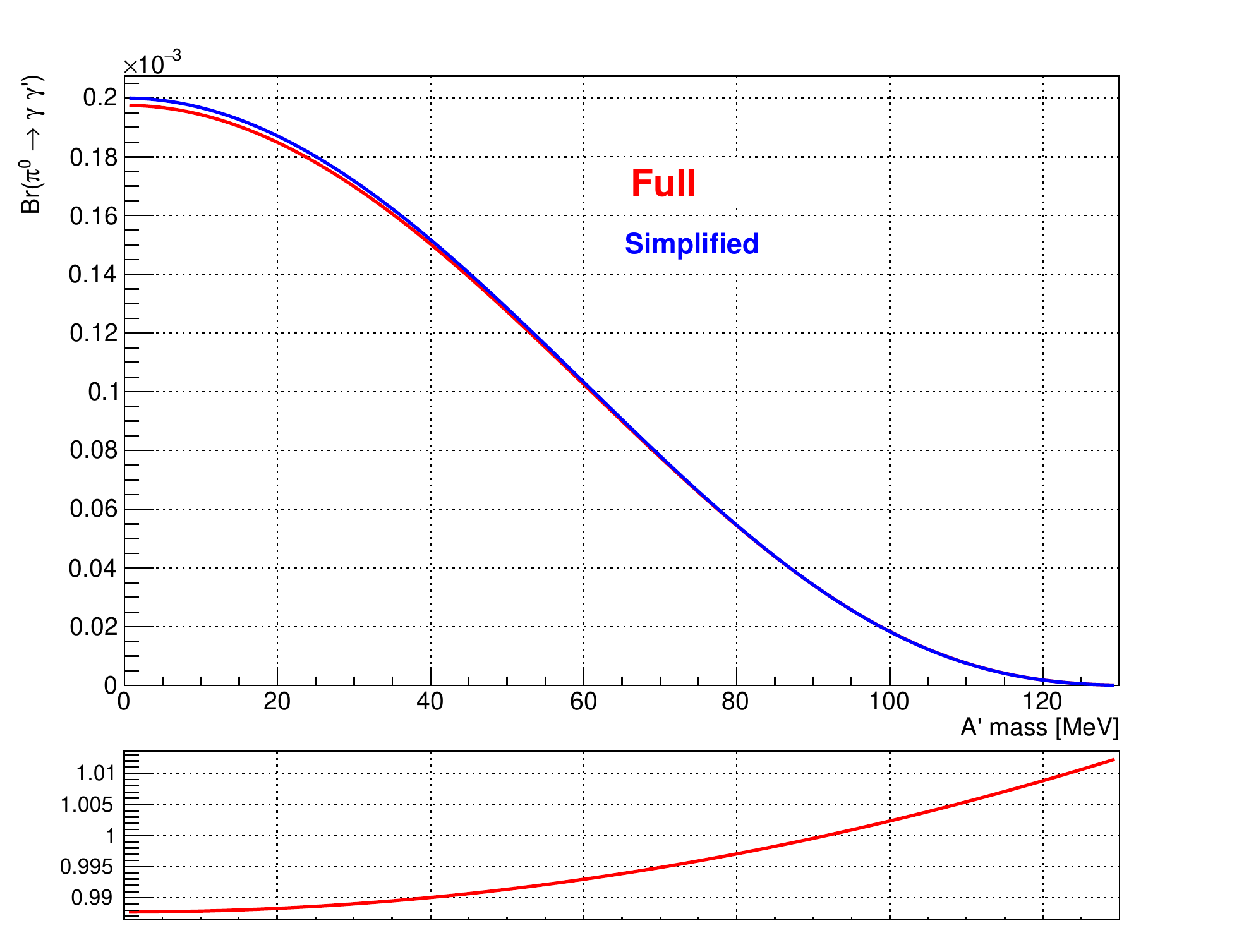}
\caption{This plot presents the branching ratio in percent as a function of the dark photon mass in MeV for $\epsilon = 0.01$. The lower panel represents the ratio between the branching fraction following from eq. (\ref{decay_width}) and the formula that Batel, Pospelov and Ritz derive in \cite{Br}, as a function of the dark photon mass in MeV.}
\label{fig:br-ggp-comp}
\end{figure}

\noindent Our treatment introduces a correction to the result obtained in \cite{Br}, since we include non-zero quark mass $m$. This allows us to correctly describe the low-energy regime (where the effect of the dark photon is sizable) of the neutral pion decay by using its decay constant.

\section{Dark sector extension}

Experimentally the end product of the above decay will be registered as a photon and a pair of an electron and a positron. We can further construct a Lorentz invariant Lagrangian containing a greater multitude of dark mesons - pseudoscalar, axial-vector, tensor and pseudo-tensor mesons. Interactions with such particles will give a contribution to the Dalitz decay width. A Lagrangian describing the interaction between a neutral pion and all of the above particles involving massive quarks will have the following form

\begin{equation}
\begin{gathered}
    \mathcal{L}=ig\overline{\Psi}\gamma^5\tau^3\Psi\pi^0-e\overline{\Psi}C\gamma_\mu A^\mu\Psi-e_1\overline{\Psi}\gamma_\mu A_1^\mu\Psi-e_2\overline{\Psi}\gamma_\mu\gamma^5A_2^\mu\Psi+ie_3\overline{\Psi}\gamma^5A_3\Psi-\\
    -ie_4\overline{\Psi}\frac{q^\mu}{|q|}\sigma_{\mu\nu}A_4^\nu\Psi+e_5\overline{\Psi}\frac{q^\mu}{|q|}\sigma_{\mu\nu}\gamma^5A_5^\mu\Psi-m\overline{\Psi}\Psi,
    \end{gathered}
\label{eq:gen-lagr}
\end{equation}

\noindent where $A_2$ is an axial-vector, $A_3$ is a pseudoscalar, $A_4$ is a tensor and $A_5$ is a pseudotensor field with their corresponding strengths of interaction. It can be shown that $\pi^0$ cannot decay into a photon and a pseudoscalar or an axial-vector particle, therefore the only particles that might influence the decay we are interested in are vector, tensor and pseudo-tensor mesons. The Feynman rules for these interactions are shown in the figure below:

\begin{figure}[H]
\centering
\includegraphics[width = 0.65\textwidth]{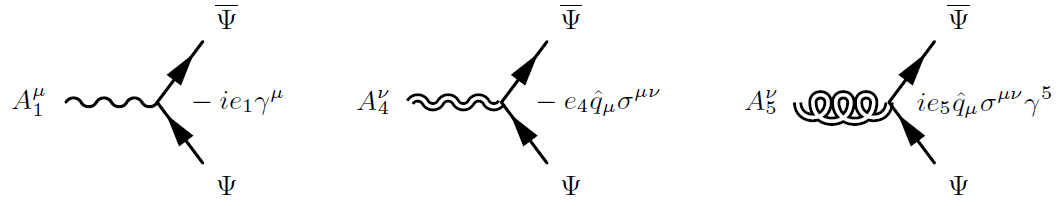}
\caption{Basic Feynman rules for the dark photon and the tensor meson fields. We have introduced the notation $\hat{q}^\mu=\frac{q^\mu}{|q|}.$}
 \label{FRules3}
\end{figure}

Once we have introduced the tensor fields $A_4^\mu$ and $A_5^\mu$ we can investigate possible decays of the neutral pion into these dark mesons. For the process $\pi^0\rightarrow\gamma A_4^\mu$ we have the following diagrams:

\vspace{5mm}

\begin{figure}[H]
\centering
\includegraphics[width = 0.46\textwidth]{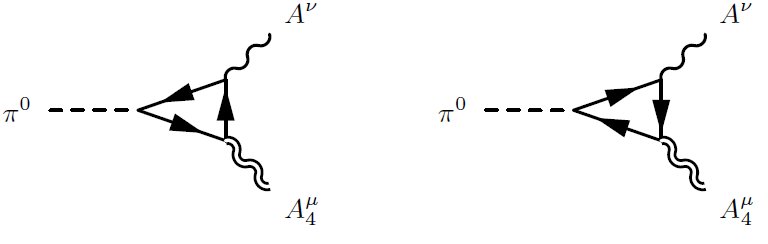}
\caption{Feynman diagram for the decay of a neutral pion into a photon and a tensor meson.}
\end{figure}

\noindent The matrix element of this process is

\begin{equation}
\begin{gathered}
    \mathcal{M}_{04\gamma}^{\mu\nu}=\frac{3}{2}ige\frac{e_4}{|k|}\int\frac{d^4p}{(2\pi)^4}\frac{\text{Tr}\left[\gamma^5(\hat{p}-\hat{q}+m)\gamma^\mu(\hat{p}-\hat{k}+m)\sigma^{\nu k}(\hat{p}+m)\right]}{(p^2-m^2)\left[(p-k)^2-m^2\right]\left[(p-q)^2-m^2\right]}+\\
    +\frac{3}{2}ige\frac{e_4}{|k|}\int\frac{d^4p}{(2\pi)^4}\frac{\text{Tr}\left[\gamma^5(\hat{p}+m)\sigma^{\nu k}(\hat{p}+\hat{k}+m)\gamma^\mu(\hat{p}+\hat{q}+m)\right]}{(p^2-m^2)\left[(p+k)^2-m^2\right]\left[(p+q)^2-m^2\right]},
    \end{gathered}
\end{equation}

\noindent where $k$ is the momentum of $A_4^\mu$. In a similar way we calculate the process $\pi^0\rightarrow\gamma A_5^\mu$. The diagrams are

\vspace{5mm}

\begin{figure}[H]
\centering
\includegraphics[width = 0.46\textwidth]{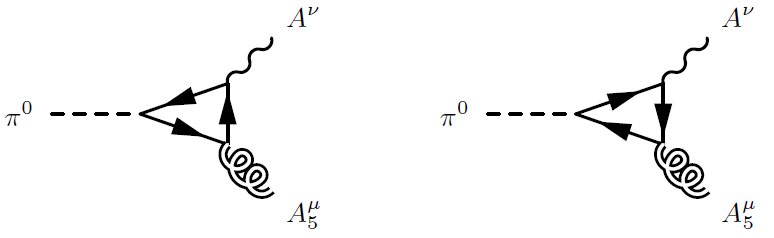}
\caption{Feynman diagram for the decay of a neutral pion into a photon and a pseudo-tensor meson.}
\end{figure}

\begin{equation}
\begin{gathered}
    \mathcal{M}_{05\gamma}^{\mu\nu}=\frac{3}{2}ge\frac{e_5}{|k|}\int\frac{d^4p}{(2\pi)^4}\frac{\text{Tr}\left[\gamma^5(\hat{p}+m)\gamma^\mu(\hat{p}+\hat{k}+m)\sigma^{\nu k}\gamma^5(\hat{p}+\hat{q}+m)\right]}{(p^2-m^2)\left[(p-k)^2-m^2\right]\left[(p-q)^2-m^2\right]}-\\
-\frac{3}{2}ge\frac{e_5}{|k|}\int\frac{d^4p}{(2\pi)^4}\frac{\text{Tr}\left[\gamma^5(\hat{p}+\hat{q}-m)\sigma^{\nu k}\gamma^5(\hat{p}+\hat{k}-m)\gamma^\mu(\hat{p}-m)\right]}{(p^2-m^2)\left[(p-k)^2-m^2\right]\left[(p-q)^2-m^2\right]}
\end{gathered}
\end{equation}

In $\mathcal{M}_{05\gamma}^{\mu\nu}$ the additional $\gamma^5$ makes the expression algebraically more complex but the physical outcome is the same as for $\mathcal{M}_{04\gamma}^{\mu\nu}$ to 2\%. Here we present the more illustrative formula for the decay width

    \begin{multline}
    \Gamma_{T\gamma}  =\\
    = \frac{4(m_\pi^2-M^2)}{\pi m_\pi^3M^2}\left(\frac{m}{f_\pi}e'e\right)^2\Bigg[\frac{\sqrt{M^2-4m^2}(M^2-m_\pi^2)\text{Log}\left(\frac{2m^2-M^2+\sqrt{M^2(M^2-4m^2)}}{2m^2}\right)}{M}+\\
    \sqrt{m_\pi^2(m_\pi^2-4m^2)}\text{Log}\left[\frac{2m^2-m_\pi^2+\sqrt{m_\pi^2(m_\pi^2-4m^2)}}{2m^2}\right]+\\
    M^2\left(2+\frac{1}{16\pi^2}\left(\text{Log}\left[1+\frac{\Lambda^2}{m^2}\right]-\frac{\Lambda^2}{\Lambda^2+m^2}\right)\right)\Bigg]^2,    
    \end{multline}

\noindent where $M$ is the mass of the field $A_4^\mu$ or $A_5^\mu$ ($M = m_{4,5}$)  and $e'$ is the corresponding interaction constant ($e'=e_{4,5}$). Since this diagram is divergent, we introduce a cut-off, $\Lambda$, which we choose to be the constituent quark mass $m$. For $M < m$ and $m_\pi < m$ we have

\begin{equation}
    \begin{gathered}
    \Gamma_{T\gamma}=\frac{4(m_\pi^2-M^2)}{\pi m_\pi^3M^2}\left(\frac{m}{f_\pi}e'e\right)^2
    \Bigg[
    \frac{\sqrt{M^2( 4m^2 - M^2 )} (m_\pi^2 -M^2)\text{arctan}\left(\frac{\sqrt{M^2(4m^2 - M^2)}}{2m^2-M^2}\right)}{M^2}-\\
    \sqrt{m_\pi^2(4m^2-m_\pi^2)}\text{arctan}\left[\frac{\sqrt{m_\pi^2(4m^2-m_\pi^2)}}{2m^2-m_\pi^2}\right]+\\
    M^2\left(2+\frac{1}{16\pi^2}\left(\text{Log}\left[1+\frac{\Lambda^2}{m^2}\right]-\frac{\Lambda^2}{\Lambda^2+m^2}\right)\right)\Bigg]^2.
    \end{gathered}
\end{equation}

\begin{figure}[H]
\centering
\includegraphics[scale=0.6]{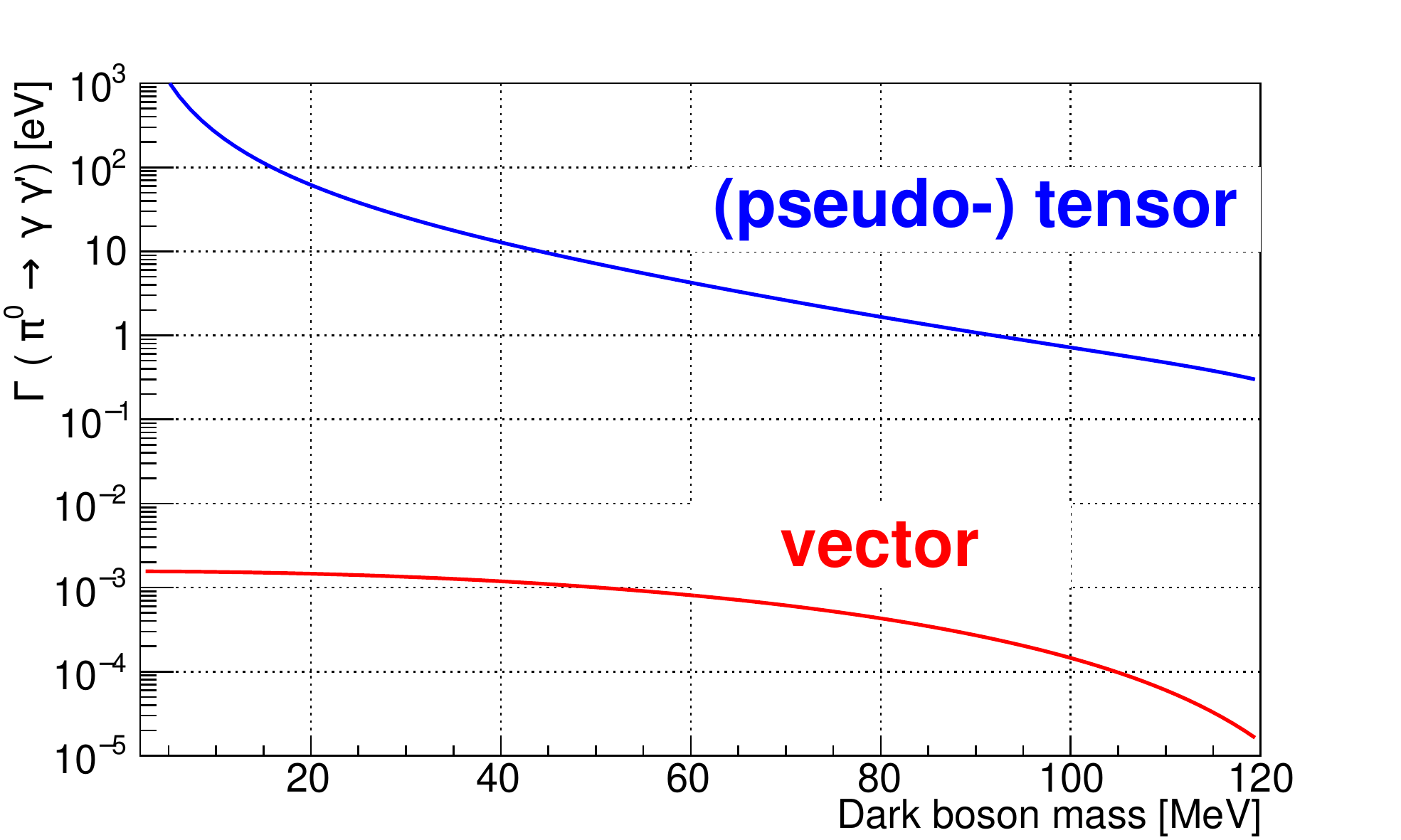}
\caption{Dependence of the decay widths $\pi^0\to\gamma\gamma'$ and $\pi^0\to\gamma A_4$ as a function of the dark boson mass. The parameter $\epsilon$ is fixed to 0.01 for both cases.}
\label{fig:pi0-dec-widths}
\end{figure}

To be able to calculate the influence of the tensor currents we investigate the decay widths of the dark mesons into pairs of electron and a positron. All these decay widths are calculated below.

\begin{figure}[H]
\centering
\includegraphics[width = 0.65\textwidth]{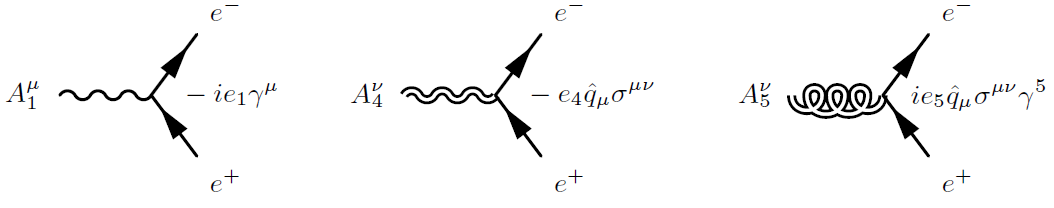}
\caption{Basic Feynman rules for the dark photon and the tensor meson fields. We have introduced the notation $\hat{q}^\mu=\frac{q^\mu}{|q|}.$}
\end{figure}

\noindent For the first process the square of the matrix element is

\begin{equation}
    |\mathcal{M}_{1e^+e^-}|^2=e_1^2\text{Tr}\left[(\hat{p}_1+m)\gamma_\mu\frac{1+\gamma^5}{2}(\hat{p}_2+m)\gamma^\mu\frac{1+\gamma^5}{2}\right]=4e_1^2(p_1p_2),
\end{equation}

\noindent where $p_1$ and $p_2$ are the positron and electron momenta, respectively. The general formula for the decay width is

\begin{equation}
    \Gamma_{1e^+e^-}=\frac{1}{8\pi}\frac{1}{2J_i+1}|\mathcal{M}_{1e^+e^-}|^2\frac{|\vec{p_1}|}{m_1^2},
\end{equation}

\noindent where $J_i$ is the angular momentum of the parent particle. Then, the decay width of the dark photon into an electron-positron pair is \cite{venelin}

\begin{equation}
    \Gamma_{1e^+e^-}=\frac{1}{3}\alpha\epsilon^2 m_1\left(1+\frac{2m_e^2}{m_1^2}\right)\sqrt{1-\frac{4m_e^2}{m_1^2}},
\end{equation}

\noindent where $\alpha=\frac{e^2}{4\pi}$ is the fine-structure constant. Applying the same treatment for the decay of the tensor mesons we obtain

\begin{equation}
    \Gamma_{4e^+e^-}=\frac{\epsilon^2\alpha}{2\pi^2}\frac{m_e^2\sqrt{m_4^2-4m_e^2}}{m_4^2}~\text{and}~\Gamma_{5e^+e^-}=\frac{\epsilon^2\alpha}{2\pi^2}\frac{m_e^2\sqrt{m_5^2-4m_e^2}}{m_5^2}.
\end{equation}

\begin{figure}[H]
\centering
\includegraphics[scale=0.5]{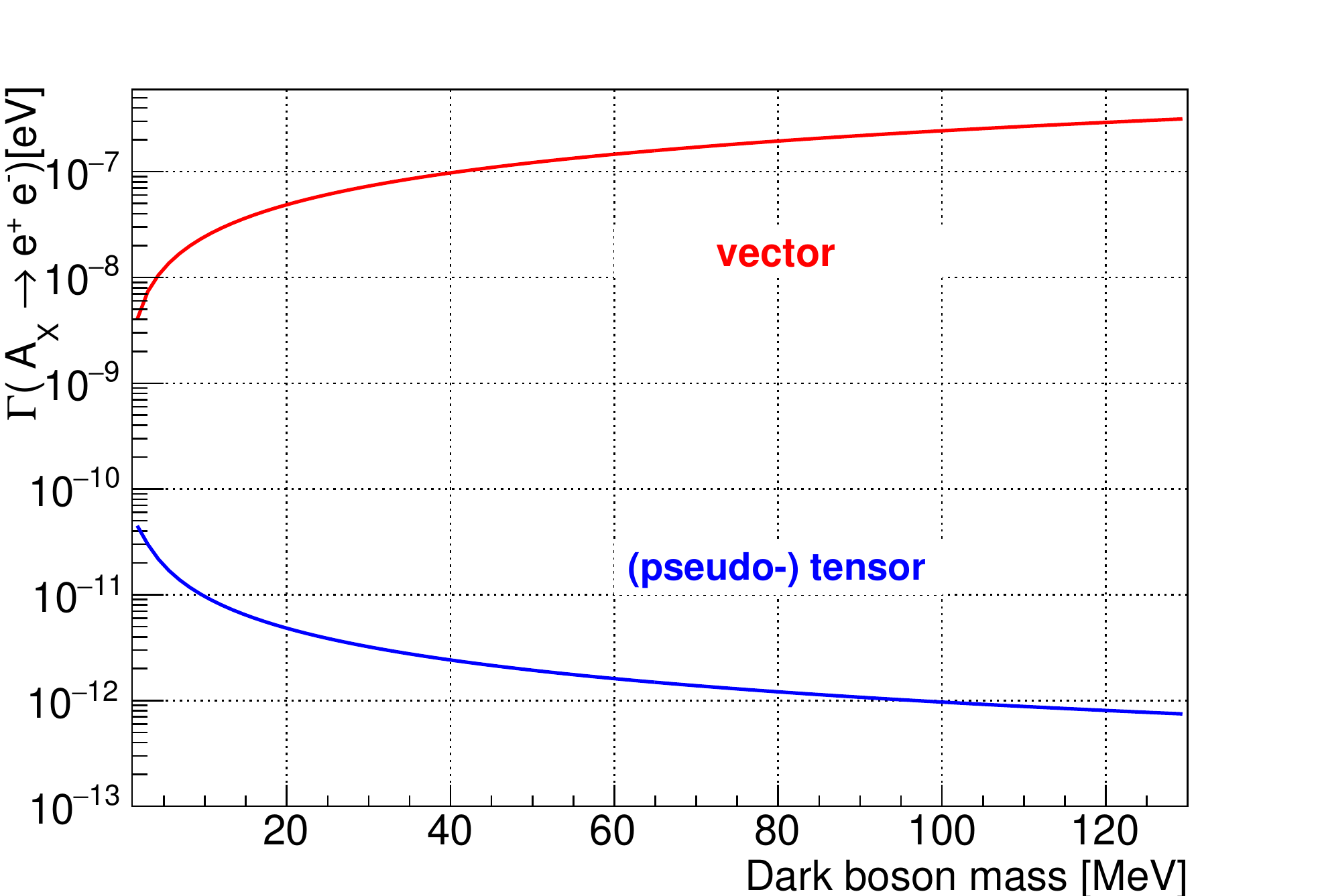}
\caption{Comparison of the decay rates of $A_1$, $A_{4,5}$ to $e^+e^-$  as a function of the dark boson mass.}
\label{fig:g-a-ee}
\end{figure}

The decay rates of $A_1$, $A_{4,5}$ to $e^+e^-$ are shown in fig. \ref{fig:g-a-ee}. 
As it can be seen, the tensor and the pseudotensor interaction term lead to a 
decay rate that decreases with the increase of the mass of the dark boson. 
The $m_e^2$ suppression is similar to the helicity suppression and the particular 
tensor form of the interaction may manifest itself in various processes 
as a contribution which violates the lepton universality.

\section{Implication on the searches for new light dark bosons}

In the context of the model developed in this paper we can explore the dependence between the dark meson mass and the interaction constant strength. 

The total contribution $\Gamma_{\text{dark}}$ of the dark particles to the measured partial decay width $\Gamma_{\text{SM}} + \Gamma_{\text{dark}}$ of the process $\pi^0\rightarrow\gamma e^+e^-$ is the sum 
\begin{equation}
\Gamma_{\text{dark}}=\Gamma_{\gamma\gamma'}*Br(A_1\to e^+e^-) + \Gamma_{\gamma A_4}*Br(A_4\to e^+e^-)+
\Gamma_{\gamma A_5}*Br(A_5\to e^+e^-).
\label{eq:full-width-pidalitz}
\end{equation}
Assuming that no new light particle interacting with $A_i$ exists, the corresponding branching fractions 
for $M_{A_i}$ are equal to 1 and the total width is given by

\begin{equation}
\Gamma_{\text{dark}}=\Gamma_{A_1} + \Gamma_{A_4}+\Gamma_{A_5}.
\label{eq:full-width-pidalitz-br-1}
\end{equation}
Then for each of the contributions we have $ \Gamma_{A_1}\leq \Gamma_{\text{dark}}$. 

The maximal magnitude of $\Gamma_{\text{dark}}$ can be inferred from the uncertainty of the 
rate of the $\pi^0\to\gamma e^+ e^-$ decay, assuming that the central value of 
$\Gamma(\pi^0\to\gamma e^+ e^-)$ is entirely determined by the Standard Model.





\begin{figure}[H]
\centering
\includegraphics[scale=0.5]{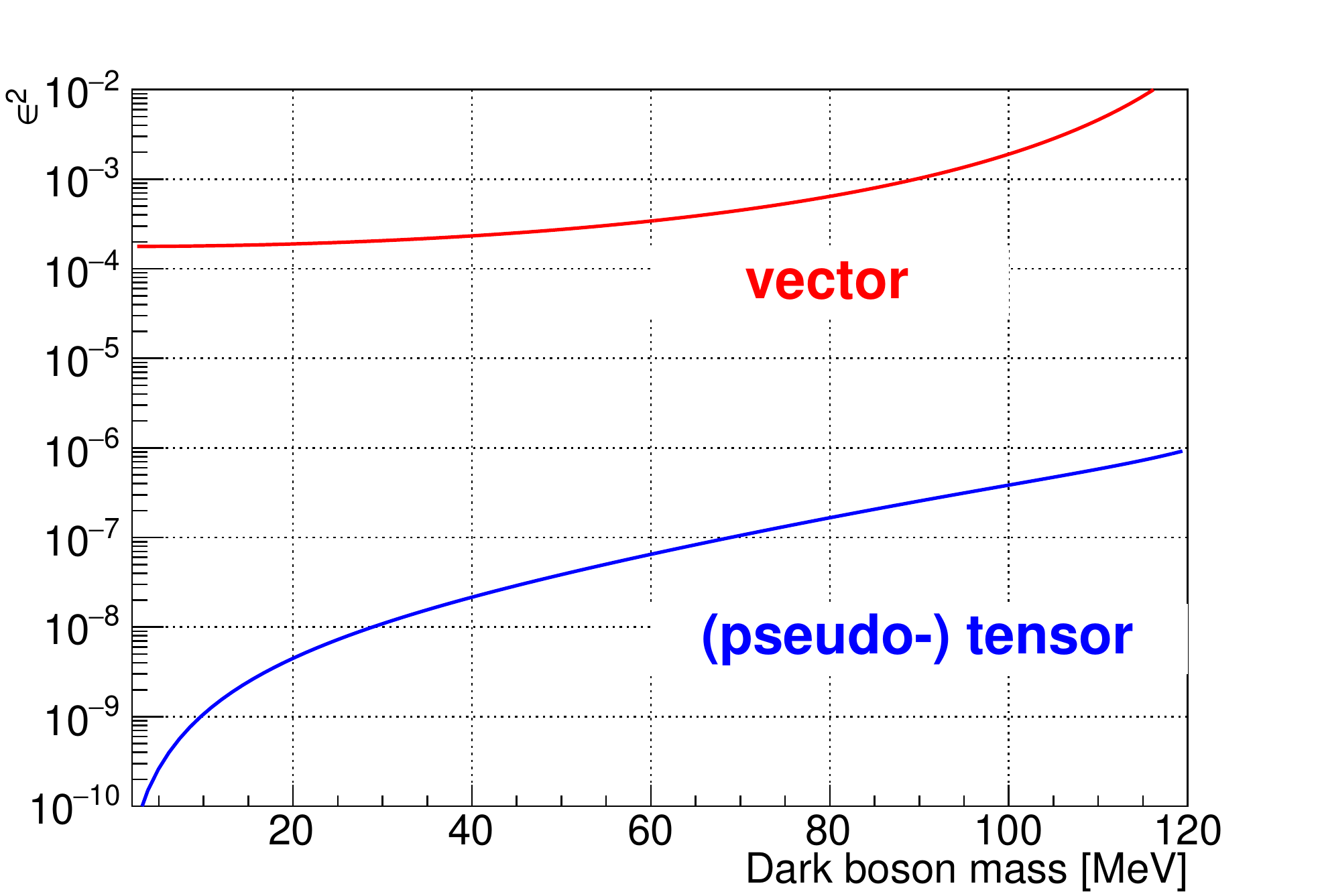}
\caption{Limits on the parameters of the dark bosons $A_1$, $A_{4,5}$ from total rate of $\pi^0\to\gamma e^+ e^-$ decay. The region above the corresponding lines can be considered excluded.}
\label{fig:limits}
\end{figure}

The uncertainty of $Br(\pi^0\to\gamma e^+ e^-)$ is estimated to be $\delta Br \approx 0.035\% $ \cite{pdg}, 
which gives an uncertainty of the decay rate $\delta \Gamma (\pi^0\to\gamma e^+ e^-)= 2.76$~meV. 
Taking  $\Gamma_{\text{dark}} \simeq \delta \Gamma (\pi^0\to\gamma e^+ e^-) $, the limits (within $1\sigma$) 
on the parameters of the dark bosons are shown in fig. \ref{fig:limits}. The limits are much stronger 
in the case of tensor interactions due to the higher  $\pi^0\to\gamma A_{4,5}$ decay rate for a 
given value of $M$ and $\epsilon$. 

\noindent In this sense, for $M=17$ MeV, the maximal allowed value of the mixing parameter for $A_4$ and $A_5$ is $\epsilon^2=3\times10^{-9}$.



\begin{figure}[H]
\centering
\includegraphics[width = 0.46\textwidth]{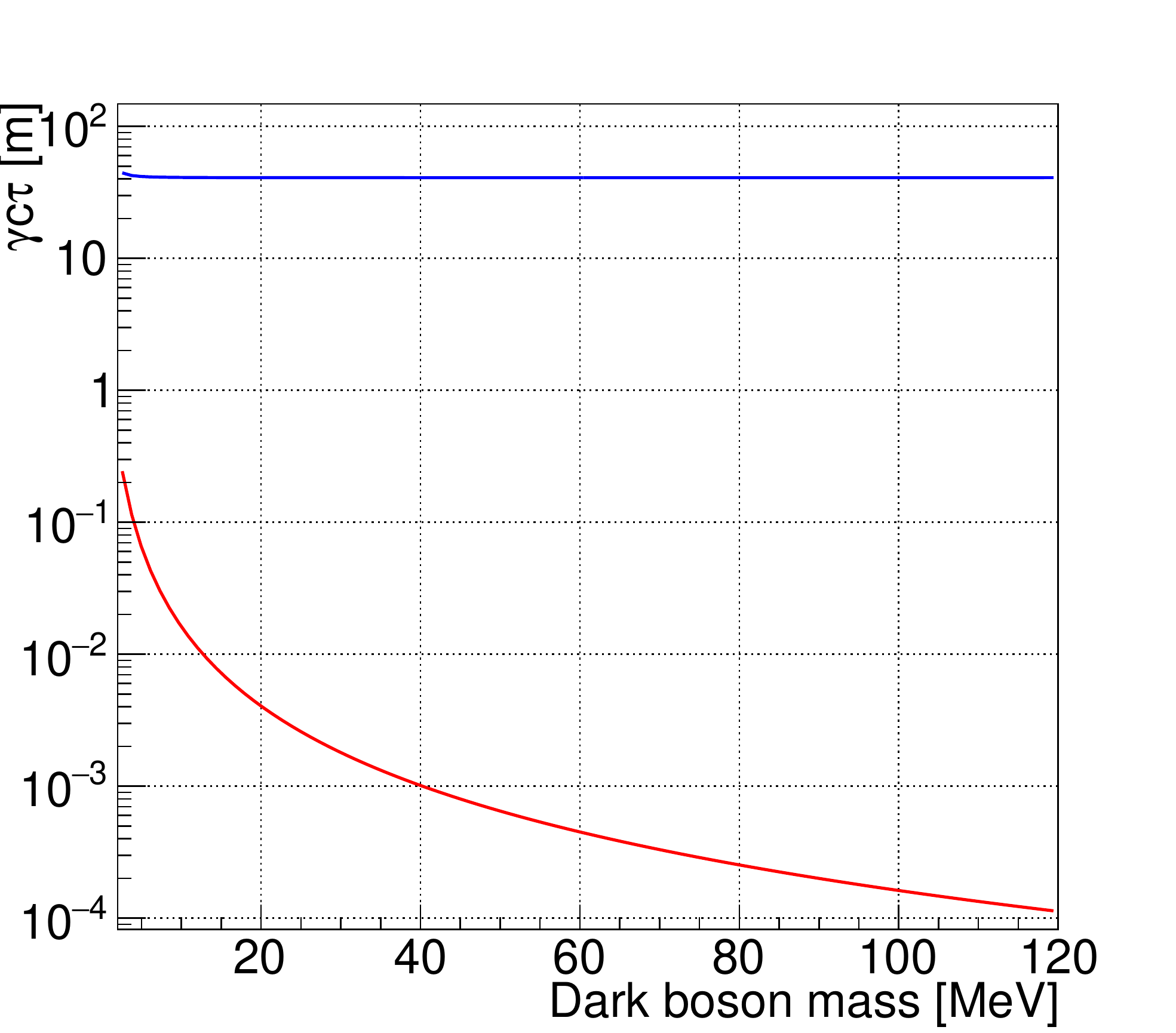}
\includegraphics[width = 0.44\textwidth]{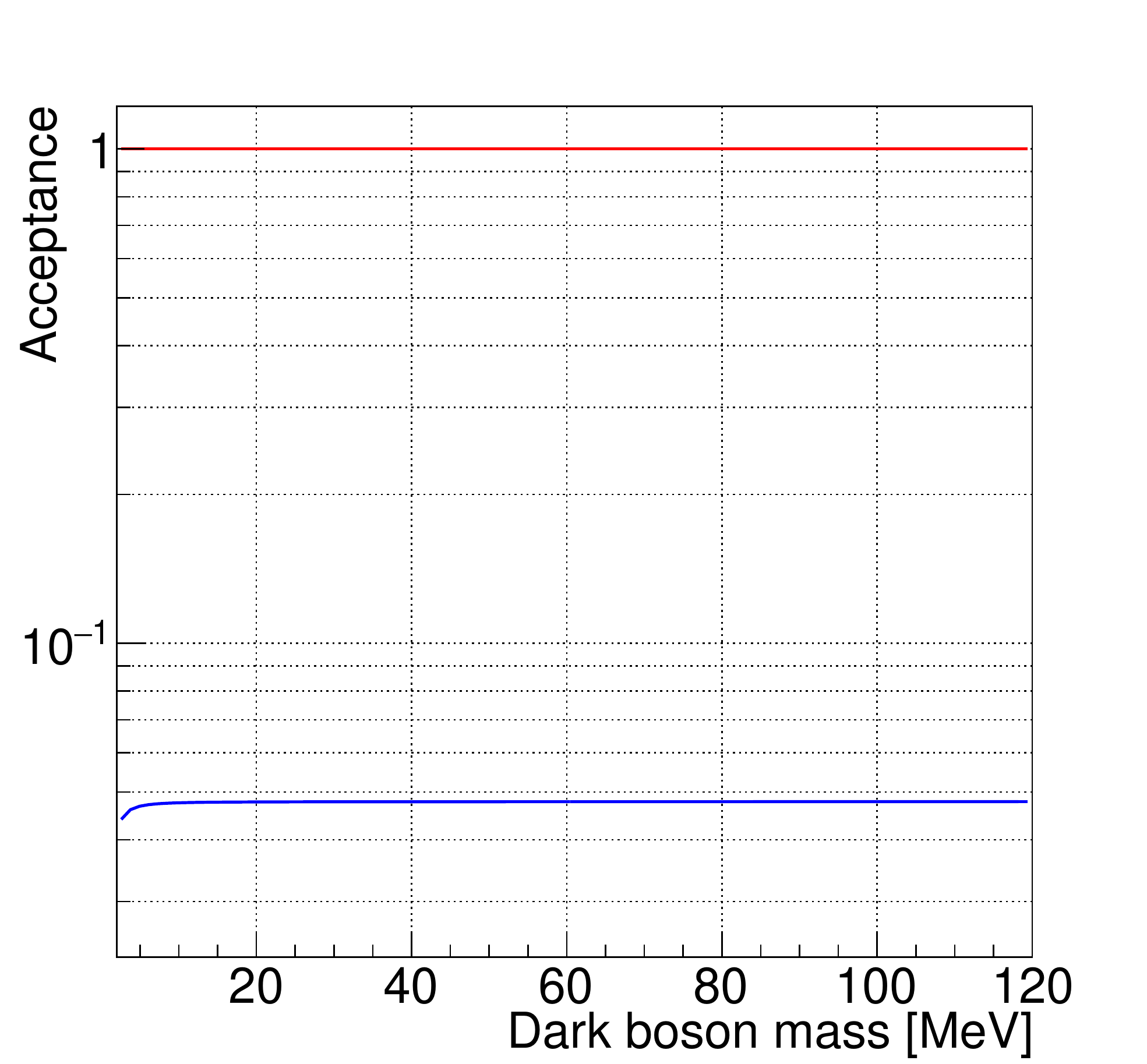}

\caption{Path length of a dark boson (red - vector, blue - (pseudo-)tensor) with energy $E = 20$~GeV as a function of a dark meson mass and 
the fraction of decays within the first 2 meters from the production point.}
\label{fig:exp-signatures}
\end{figure}

The value of $\Gamma_{\text{dark}}$ can also be studied in the exclusive reconstruction of the final state,
where the electron-positron pair is used to determine the decay vertex of $\pi^0$. 
However, the dark photon $A_1$ and the tensor interacting $A_4$ and $A_5$ have different decay rates
which leads to different lifetimes and correspondingly different distribution of the decay vertex of the dark boson. 
For example, for dark bosons with energy $E=20$~GeV the dependence of the $\gamma c \tau$ with 
the mass of the particle (for $\epsilon = 0.001$) is shown in fig. \ref{fig:exp-signatures}, left. 
It should be noted that $\gamma c \tau = const$ for tensor interaction since the 
decay rate is proportional to $1/m_{4,5}$ and $\tau \sim m_{4,5}$, while $\gamma = E/m_{4,5}$. 
Assuming that it is possible to reconstruct the event only if the
$A_{4,5}$ decay vertex is within few meters ($\Delta L$) from the $\pi^0$ decay vertex (usually the case 
for most fixed target experiments, eg. \cite{NA48/2}),
the experimental acceptance is shown in fig. \ref{fig:exp-signatures}, right. 
In this particular case $\Delta L = 2$~m was used and a further assumption was made 
that all $A_{4,5}$ decay products are traversing the experimental setup, i.e. 100 \% geometrical efficiency. 
We do not take into account the efficiency in reconstructing the final state
since it is experimental specific and should be subject to detailed experimental study. 
In real experiments, both the geometrical and the reconstruction efficiencies will further reduce the events acceptance.

It should be noted, however, that such considerations are usually not enough to extract the 
correct information from a given experimental result to a different model. 
Many effects have to be considered when performing such procedure, 
including proper simulation of the experimental reconstruction efficiency, 
different background topology, the underlying assumption, etc. 
Such detailed analysis with the full understanding of these effects can only 
be performed within the respected collaborations.

\section{Conclusions}

The proposed Lagrangian in (\ref{eq:gen-lagr}) describes in the most general form 
the interaction of spin 0 and spin 1 particle with the Standard Model fermions. 
We applied this model to the decay of $\pi^0$ and obtained the 
relevant contributions to the decay width. 


So far no tensor current interactions have been considered, 
however in the general treatment 
they should not be excluded since they introduce a sizeable effect. 
The decay width of dark vector mesons interacting through tensor currents 
is dependent on the mass of the lepton in the final state 
inducing a lepton universality violation effect. 
This topic is left for further investigation together with the influence of tensor interacting bosons 
on the muon magnetic moment, a subject to a subsequent work. 
This model can be probed in all future experiments where neutral pions are abundantly produced
in events, allowing their inclusive reconstruction.

\section*{Acknowledgements} The authors would like to thank Enrico Nardi from LNF-INFN for the 
fruitful discussions. 
MN was partially supported within the program "Young scientists and postdocs", RD22-2579/01/04/2021, 
and VK acknowledges partial support from BNSF KP-06-D002\_4/15.12.2020
within MUCCA, CHIST-ERA-19-XAI-009.


\begin{thebibliography}{9}
\bibitem{holdom}
B. Holdom,
\textit{Two U(1)'s and epsilon charge shifts}, Phys. Lett. B \textbf{166}, 196-198, 1986.
\bibitem{rev1}
A. Filippi and M. De Napoli.
\textit{Searching in the dark: the hunt for the dark photon
}, Phys. Rev. \textbf{5}, 100042, 2020.
\bibitem{rev2}
A. Boveia and C. Doglioni.
\textit{Dark matter searches at colliders}, Ann. Rev. Nucl. Part. Sci. \textbf{68}, 429-459, 2018.
\bibitem{rev3}
B. Penning.
\textit{The pursuit of dark matter at colliders—an overview
}, J. Phys. G \textbf{45}, 063001, 2018.
\bibitem{Be}
A. J. Krasznahorkay, M. Csatlós, L. Csige, Z. Gácsi, J. Gulyás, M. Hunyadi, I. Kuti, B. Nyakó, L. Stuhl, J. Timár, T. Tornyi, Zs. Vajta, T. Ketel and A. Krasznahorkay.
\textit{Observation of anomalous internal pair creation in $^8$Be: a possible indication of a light, neutral boson}, Phys. Rev. Lett. \textbf{116}, 042501, 2016.
\bibitem{g2}
B. Abi et. al.
\textit{Measurement of the positive muon aomalous magnetic moment to 0.46 ppm}, Phys. Rev. Lett. \textbf{126}, 141801, 2021.
\bibitem{SM1}
A. Aleksejevs, S. Barkanova, Yu.G. Kolomensky and B. Sheff.
\textit{A Standard Model Explanation for the "ATOMKI Anomaly"}, arXiv:2102.01127 [hep-ph].
\bibitem{SM2}
Borsanyi, S., Fodor, Z., Guenther, J.N. et al. 
\textit{Leading hadronic contribution to the muon magnetic moment from lattice QCD},  Nature 593, 51–55, 2021.
\bibitem{leptoquarks}
K. Cheung.
\textit{Muon anomalous magnetic moment and leptoquark solutions
}, Phys. Rev. D \textbf{64}, 033001, 2001.
\bibitem{supersymmetry}
M. Davier and W. Marciano.
\textit{The theoretical prediction for the muon anomalous magnetic moment}
\bibitem{pdg}
K.A. Olive et al. (Particle Data Group), Chin. Phys. C \textbf{38}, 090001, 2014.
\bibitem{photon1}
M. Hostert, K. Kaneta and M. Pospelov.
\textit{Pair production of dark particles in meson decays}, Phys. Rev. D \textbf{102}, 055016, 2020.
\bibitem{photon2}
NA48/2 Collaboration.
\textit{Search for the dark photon in $\pi^0$ decays}, Phys. Lett. B \textbf{746}, 178-186, 2015.
\bibitem{photon3}
SHiP Collaboration.
\textit{Sensitivity of the SHiP experiment to dark photons decaying to a pair of charged particles}, arXiv:2011.05115v2 [hep-ex].
\bibitem{photon4}
Y. Zhang and Y. Zhao.
\textit{Unconventional dark matter models: a brief review}, Sci. Bull. \textbf{60}, 986-994, 2015.
\bibitem{chizhov1}
M. Chizhov.
\textit{Vector meson couplings to vector and tensor currents in extended NJL quark model}, JETP Letters \textbf{80}, 73–77, 2004.
\bibitem{chizhov_naydenov}
M. Chizhov and M. Naydenov
\textit{Isospin-invariant Nambu – Jona-Lasinio model with complete set of spin-1 excitations}, AIP Conference Proceedings, \textbf{2075}, 090025, 2019.
\bibitem{osipov}
A. Osipov and M. Volkov.
\textit{Chiral transformations of spin-1 mesons in the non-symmetric vacuum
}, Annals Phys. \textbf{382}, 50–63, 2017.
\bibitem{Eguchi}
T. Eguchi.
\textit{New approach to collective phenomena in superconductivity models}, Phys. Rev. D \textbf{14}, 2755–2763, 1976.
\bibitem{Griffiths}
D. Griffiths. \textit{Introduction to elementary particles}. WILEY-VCH. pp. 135, 2008.
\bibitem{pdg2}
P.A. Zyla et al. (Particle Data Group).
\textit{Review of particle physics}, Prog. Theor. Exp. Phys. \textbf{2020}, 083C01, 2020.
\bibitem{decay_constant}
C. Qin, L. Xiao-Nan, L. Xin-Qiang and S. Fang.
\textit{Decay constants of pseudoscalar and vector mesons with improved holographic wavefunction}, Chinese Phys. C \textbf{42}, 073102, 2018
\bibitem{DP_Production}
E. Nardi, C. Carvajal, A. Ghoshal, D. Meloni and M. Raggi.
\textit{Resonant production of dark photons in positron beam dump experiments}, Phys. Rev. D \textbf{97}, 095004, 2018.
\bibitem{meson_decays}
J. Bijnens and K. Kampf.
\textit{Neutral pseudoscalar meson decays: $\pi^0\rightarrow\gamma\gamma$ and $\eta\rightarrow\gamma\gamma$ in $SU(3)$ limit}, Nucl. Phys. B Proc.Suppl. 207-208, 220-223, 2010.
\bibitem{Br}
B. Batell, M. Pospelov and A. Ritz.
\textit{Exploring portals to a hidden sector through fixed targets}, Phys. Rev. D, \textbf{80}, 095024, 2009.
\bibitem{weinberg}
S. Weinberg.
\textit{Why do quarks behave like bare Dirac particles?}, Phys. Rev. Lett., \textbf{65}, 1181-1193, 1990.
\bibitem{venelin}
M. Raggi and V. Kozhuharov.
\textit{Proposal to search for a dark photon in positron on
target collisions at DA$\Phi$NE linac}, Adv. High Energy Phys., \textbf{2014}, 959802, 2014.
\bibitem{NA48/2}
J.~R.~Batley \textit{et al.} [NA48/2],
\textit{Search for the dark photon in $\pi^0$ decays},
Phys. Lett. B \textbf{746} (2015), 178-185.
\end{thebibliography}
\end{document}